\documentclass[iop]{emulateapj}
\usepackage{apjfonts}
\usepackage{graphicx}
\usepackage{amsmath}
\usepackage{threeparttable}
\usepackage[utf8]{inputenc}
\usepackage{color}
\usepackage{natbib}
\newcommand\bgin[1]{\begin{#1}}
\newcommand{\bim}{\begin{itemize}}
\newcommand{\eim}{\end{itemize}}
\newcommand{\beq}{\bgin{equation}}
\newcommand{\eeq}{\end{equation}}

\newcommand{\msol}{h^{-1}M_{\Sun}}
\newcommand{\Msol}{M_{\Sun}}

\newcommand{\Mpc}{\mathrm{Mpc}}
\newcommand{\dd}{\mathrm{d}}
\def\epsdm{\epsilon_{\rm DM}}
\def\epsfb{\epsilon_{\rm f}}

\def\fgasfh{f_{\mathrm{gas,500}}}

\def\mgas{M_{\mathrm{gas}}}
\def\mgasfh{M_{\mathrm{gas,500}}}

\def\xbr{x_{\mathrm{break}}}

\begin{document}

\title{Constraints on the optical depth of galaxy groups and clusters}

\author{Samuel Flender$^{1,2}$, Daisuke Nagai$^{3}$, Michael McDonald$^{4}$}
\affil{$^1$HEP Division, Argonne National Laboratory, 9700 S. Cass Ave., Lemont, IL 60439, USA}  
\affil{$^2$Kavli Institute for Cosmological Physics, The University of Chicago, Chicago, IL 60637, USA}
\affil{$^3$Department of Physics, Yale University, New Haven, CT 06520, USA}
\affil{$^4$MIT Kavli Institute for Astrophysics and Space Research, Massachusetts Institute of Technology, \\ 77 Massachusetts Avenue, Cambridge, MA 02139, USA}

\begin{abstract}
Future data from galaxy redshift surveys, combined with high-resolutions maps of the cosmic microwave background, will enable measurements of the pairwise kinematic Sunyaev-Zel'dovich (kSZ) signal with unprecedented statistical significance. This signal probes the matter-velocity correlation function, scaled by the average optical depth ($\tau$) of the galaxy groups and clusters in the sample, and is thus of fundamental importance for cosmology. However, in order to translate pairwise kSZ measurements into cosmological constraints, external constraints on $\tau$ are necessary. In this work, we present a new model for the intra-cluster medium, which takes into account star-formation, feedback, non-thermal pressure, \emph{and} gas cooling. Our semi-analytic model is computationally efficient and can reproduce results of recent hydrodynamical simulations of galaxy cluster formation. 
We calibrate the free parameters in the model using recent X-ray measurements of gas density profiles of clusters, and gas masses of groups and clusters. 
Our observationally calibrated model predicts the average $\tau_{500}$ (i.e.\ the integrated $\tau$ within a disk of size $R_{500}$) to better than 6\% modeling uncertainty (at 95\% confidence level). If the remaining uncertainties associated with other astrophysical uncertainties and X-ray selection effects can be better understood, our model for the optical depth should break the degeneracy between optical depth and cluster velocity in the analysis of future pairwise kSZ measurements and improve cosmological constrains from the combination of upcoming galaxy and CMB surveys, including the nature of dark energy, modified gravity, and neutrino mass.

\end{abstract}

\section{Introduction}
Clusters of galaxies are the most massive gravitationally collapsed objects in the Universe today, containing hundreds of galaxies. On scales of up to a few hundred $\Mpc$, clusters move on average toward each other due to their mutual gravitational attraction. A measurement of this long-range pairwise motion has the potential to shed new insights into dynamical dark energy, modified gravity models, and neutrino mass. 

From the point of view of an observer, two clusters moving toward each other appear with opposite line-of-sight velocities, where the cluster that is further away has a velocity component toward the observer, and vice versa. The average velocity at which clusters at a given separation move toward each other --- i.e., the \emph{pairwise velocity} --- can thus in principle be estimated using only the information about their line-of-sight peculiar velocities \citep{Ferreira:1998id}. In practice however, peculiar velocities are difficult to measure as these measurements often require precise measurement of distances as well as redshifts.

The Sunyaev-Zel'dovich (SZ) effect \citep{Zeldovich:1969ff} refers to the inverse Compton scattering of cosmic microwave background (CMB) photons with free, high-energy electrons, and is further decomposed into the thermal SZ (tSZ) and the kinematic SZ (kSZ) components. For clusters of galaxies, the dominant component is the tSZ effect, which is sourced by the electrons that reside inside the hot intra-cluster medium (ICM). The tSZ effect creates a spectral distortion in the CMB blackbody radiation in the form of a temperature decrement (increment) at frequencies below (above) 217 GHz, and can thus be used to detect new clusters in CMB data (see, e.g., \citealt{hasselfield13}, \citealt{Bleem:2014iim}, \citealt{planck15XXVII}). The kSZ effect on the other hand is sourced by CMB photons scattering off electrons that have a non-zero peculiar velocity with respect to the CMB rest frame. The kSZ signal from clusters is thus a potential proxy for their line-of-sight peculiar velocities. Prospects for reconstructing the peculiar velocities of clusters via their kSZ signature have been discussed in e.g.,\ \citet{Sunyaev:1980nv, 1991ApJ...372...21R,2001AA...374....1A}.

The detection of the kSZ signal for an individual cluster has been limited to a handful of massive galaxy clusters (e.g., \citealt{sayer13}), because of its identical spectral dependence compared to the CMB, and its small amplitude, which is typically of the order of only a few $\mu$K. However, the pairwise motion of clusters, combined with the kSZ effect, creates a distinct pattern in the CMB, consisting of subtle temperature increments and decrements at the cluster locations, depending on their line-of-sight momenta. We call this distinct CMB pattern created by cluster pairs the \emph{pairwise kSZ signal}.

The pairwise kSZ signal has been measured using CMB data from the Atacama Cosmology Telescope (ACT) with galaxy positions from the Baryon Oscillation Spectroscopic Survey (BOSS) \citep{Hand:2012ui,DeBernardis:2016pdv}, CMB data from Planck with galaxy positions from the Sloan Digital Sky Survey (SDSS) \citep{Ade:2015lza}, and CMB data from the South Pole Telescope (SPT) with cluster positions from the Dark Energy Survey (DES) \citep{Soergel:2016mce}. In addition to these pairwise kSZ measurements, it is also possible to detect the kSZ signal by stacking CMB patches around galaxy locations weighted by the reconstructed velocity field \citep{Schaan:2015uaa}, or by cross-correlating the squared CMB temperature map with galaxy positions, as done in \citet{Hill:2016dta}, using data from Planck and the Wide-field Infrared Survey Explorer (WISE) survey.

Looking forward, \citet{Keisler:2012eg} forecast detection significances for the pairwise kSZ signal of $18\sigma-30\sigma$ with the next-generation version of SPT. \citet{Flender:2015btu} (hereafter F16)  predict that future data from the Advanced ACTPol experiment, combined with cluster catalogs from the Dark Energy Spectroscopic Instrument (DESI), can enable detection significances of $\sim20\sigma-50\sigma$, and even higher with cluster catalogs that go to masses below $M_{200}=10^{14}\msol$. \citet{Dore:2016tfs} project a detection significance of $\sim55\sigma$ with galaxy catalogs from the proposed SPHEREx experiment in combination with ACTPol CMB data. \citet{Ferraro:2016ymw} predict a detection significance of the projected squared kSZ signal with data from the WISE survey and Advanced ACTPol of $\sim120$, and even more with galaxy catalogs from the SPHEREx. \cite{Sugiyama:2016rue} predict that the future Advanced ACTPol and CMB-StageIV experiments, combined with galaxy surveys from DESI, should achieve measurements of the pairwise kSZ power spectrum with statistical significances of $10-100\sigma$. We summarize these forecasts in Table\,\ref{table_forecasts}.

\begin{table*}
{\footnotesize
\begin{center}
\begin{tabular}{|c|c|c|c|}
\hline
data scenario            & method   & predicted S/N & Reference \\ \hline \hline
SPT-3G $\times$ DES      & pairwise kSZ & $\sim18-30$         & \citet{Keisler:2012eg}   \\ \hline
Adv.ACTPol $\times$ DESI & pairwise kSZ & $\sim20-57$         & \citet{Flender:2015btu}  \\ \hline
Adv.ACTPol $\times$ SPHEREx & pairwise kSZ & $\sim 55$ & \citet{Dore:2016tfs} \\ \hline
Adv.ACTPol $\times$ WISE & projected kSZ & $\sim 120$ & \citet{Ferraro:2016ymw} \\ \hline
Adv.ACTPol $\times$ DESI & pairwise kSZ power spectrum & $\sim 30$ & \citet{Sugiyama:2016rue} \\ \hline
CMB StageIV $\times$ DESI & pairwise kSZ power spectrum & $\sim 50-100$ & \citet{Sugiyama:2016rue} \\ \hline

\end{tabular}
\end{center}
}
\caption{\label{table_forecasts} kSZ detection forecasts for future experiments from various references.}
\end{table*}

From a cosmological perspective, measurements of the kSZ effect have the potential of probing dark energy and modified gravity  \citep{Mueller:2014nsa,Kosowsky:2009nc,Bhattacharya:2007sk,Keisler:2012eg,DeDeo:2005yr,HernandezMonteagudo:2005ys,Ma:2013taq,Alonso:2016jpy}, as well as the sum of the neutrino masses \citep{Mueller:2014dba}. However, the constraining power of the kSZ signal is fundamentally limited by our understanding of the integrated electron density, i.e.,\ the \emph{optical depth}, of the galaxy clusters sourcing the signal. The optical depth depends on the properties of the halo hosting the galaxy cluster, such as its mass and concentration, as well as astrophysical effects such as star-formation and feedback from Active Galactic Nuclei (AGN) and Supernovae (SNe).

F16 demonstrated that for a fixed cluster sample, the optical depth (and thus the kSZ amplitude) varies by a factor of $\sim2$ between models with and without star-formation and feedback, i.e.,\ in the absence of any other constraints, the uncertainty is $\sim$100\%. \citet{Battaglia:2016xbi} (hereafter B16) presented the results for 3 different hydrodynamical simulations with varying input cluster astrophysics, demonstrating a simple scaling relation between the integrated $\tau$ and the halo mass. By comparing the run with AGN feedback to the run without it (but including radiative cooling and star formation), B16 reported a modeling uncertainty (associated with the AGN feedback) in the normalization of the scaling relation of 12\%. For the scaling relation between the integrated $\tau$ and the integrated Compton-$y$ parameter, this uncertainty is only 8\%. The modeling uncertainty is however much higher (around 50\%), when comparing the runs with star-formation and cooling to the non-radiative run, similar to the results of F16. Given the high significance of kSZ measurements expected with future experiments, a better understanding of the optical depth is thus crucial in order to realize the statistical power of the upcoming galaxy and CMB surveys for cosmology.

Hydrodynamical simulations are computationally expensive, such that only a small number of different ICM models can be studied. Ideally, however, we would want to create a large number of ICM models and use machine learning algorithms and observational data to solve for both cosmology and astrophysics at the same time. For parameter estimation, \emph{semi-analytical model} is a method of choice, as it allows us to study a large number of ICM models with considerably less computational cost than hydrodynamical simulations. 

In this work we constrain the optical depth profile of galaxy clusters using a semi-analytical model that is computationally efficient and has only a small number of free parameters. Our ICM model is based on the model introduced in \citet{Ostriker:2005ff}, and modified in \citet{Shaw:2010mn}. However, these models did not take into account the effects of gas cooling, which make them unable to provide a reasonable description of X-ray observations in cluster cores. Thus, in this work, we extend these ICM models to take into account the effects of gas cooling of cluster cores by introducing the effective equation of state (EOS) in the cooling region. We constrain the parameter space of that model using recent X-ray measurements of gas density profiles of clusters from \citet{McDonald:2013fka} and gas mass in groups and clusters from \citet{0004-637X-640-2-691}, \citet{Sun:2008eh} and \citet{2015A&A...573A.118L}.

The outline of this work is as follows. In Section\,2 we will introduce the kSZ effect and its cosmological implications. In Section\,3 we will describe our ICM model. In Section\,4 we will compare the $\tau-M$ relation in our model to the recent hydrodynamical simulations. In Section\,5 we will present the main results of this work, including the observationally calibrated $\tau$ profiles, along with its associated MCMC analysis. We will discuss future prospects and challenges in Section\,6. Our main results are summarized in Section\,7. Throughout this work we assume a WMAP7 cosmology \citep{2011ApJS..192...18K} with $h=0.71$, $\Omega_M = 0.26$ and $\Omega_b = 0.0448$.

\section{Cosmology with the kSZ effect}

\subsection{The kSZ effect}
The kSZ effect is caused by the inverse Compton-scattering of CMB photons with free electrons moving with non-zero peculiar velocities with respect to the CMB rest frame. Along a given line of sight, the kSZ temperature is given by the integral
\beq
\frac{\Delta T_{\mathrm{kSZ}}}{T_{\mathrm{CMB}}} = \frac{\sigma_T}{c}\int\mathrm{d}l\,n_e(l) v_{\mathrm{los}}(l),
\eeq
where $T_{\mathrm{CMB}}=2.725\,\mathrm{K}$ is the average blackbody temperature of the CMB, $\sigma_T$ is the Thomson cross section, $n_e$ is the number density of electrons along the line of sight, and $v_{\mathrm{los}}$ their peculiar velocity along the line of sight, where $v_{\mathrm{los}}>0$ for objects moving toward the observer.

For a collapsed object, e.g.,\ a cluster, all electrons bounded within the accretion shock radius, $R_{\rm shock} \simeq 4-5R_{500}$ \citep{Lau:2014lwa}, move in bulk with the halo peculiar velocity, $v_{\rm los}$ (see Sec.\,\ref{sec:residual_uncertainties} for a discussion of velocity substructures), where $R_{500}$ is defined in Eq.\,\ref{eq:Mdelta}. For such an object we can take $v_{\mathrm{los}}$ out of the integrand, and write its kSZ contribution as
\beq
\frac{\Delta T_{\mathrm{kSZ}}}{T_{\mathrm{CMB}}} = \tau \frac{v_{\mathrm{los}}}{c},
\label{eq:ksz}
\eeq
where the \emph{optical depth}, $\tau$, of the cluster at the projected angular distance, $\theta$, away from the cluster center on the sky is given by 
\beq
\tau(\theta) = 2 \sigma_T \int_{\theta d_A(z)}^{R_{\rm shock}}  \mathrm{d}r \, n_e(r)  \bigg(\frac{r}{\sqrt{r^2-\theta^2d_A^2(z)}}\bigg) \, ,
\eeq
where $d_A(z)$ is an angular diameter distance for an object at redshift $z$. 

Current high-resolution CMB experiments such as SPT and ACT have finite beam sizes of around 1\,arcmin. A more observationally relevant quantity is thus the average kSZ temperature within an aperture of size $\theta$,
\beq
\Delta T_{\mathrm{kSZ},\theta} \equiv \frac{2\pi\int_0^{\theta}d\theta^{\prime}\theta^{\prime} \Delta T_{\mathrm{kSZ}}(\theta^{\prime})}{\pi\theta^2} \,.
\eeq
Similarly, we define the integrated optical depth within the aperture $\theta$,
\beq
\tau_{\theta}\equiv\frac{2\pi\int_0^{\theta}d\theta^{\prime}\theta^{\prime}\tau(\theta^{\prime})}{\pi\theta^2} \,.
\label{eq:tautheta}
\eeq
If the aperture size $\theta$ is chosen to match the angular extent of the cluster in the sky, it follows from Eq.\,\ref{eq:ksz} that the integrated kSZ signal can be related to the integrated optical depth as 
\beq
\frac{\Delta T_{\mathrm{kSZ},\theta}}{T_{\mathrm{CMB}}} = \tau_{\theta} \frac{v_{\mathrm{los}}}{c} .
\eeq

The integrated kSZ temperature within the aperture $\theta$ of a cluster receives not only the kSZ contribution from that cluster, but also from all objects along the same line of sight. However, these additional kSZ contributions add noise to the pairwise kSZ measurement discussed below, but not a bias, because the velocities are distributed symmetrically around zero, on average. Using lightcone simulations from \citet{Flender:2015btu}, \citet{Soergel:2016mce} estimate that noise level to be around $\sim7\%$, but it is expected to be much less with larger sky coverage (e.g.,\ $\lesssim2\%$ with ACTPol$\times$DESI). Here, we assume that the aperture size $\theta$ is chosen to match the cluster scale. However, if $\theta$ is chosen to be much larger, then the integrated kSZ signal could receive contributions from objects located close to the cluster with correlated velocities \citep{Schaan:2015uaa}.

\subsection{The pairwise kSZ signal}

Due to their mutual gravitational attraction, clusters of galaxies move on average toward each other. If we had the velocity vector for each cluster in a given sample, then we could compute the \emph{pairwise velocity}, 
\beq
v_{12}(r,z)=\langle \mathbf{v}_1 - \mathbf{v}_2 \rangle_r ,
\eeq
where the brackets denote the average at comoving separation $r$. For instance, at a separation of $r=100$\,Mpc and at $z=0.5$, clusters move on average toward each other with a velocity of order $100\,\mathrm{km}/\mathrm{s}$. Due to the kSZ effect, this pairwise motion imprints a temperature signal into the CMB, which we call the \emph{pairwise kSZ signal}. Analogous to Eq.\,\ref{eq:ksz}, the pairwise kSZ signal is given by the pairwise velocity times the average optical depth of the cluster sample,
\beq
\frac{\Delta T_{\mathrm{pkSZ}}}{T_{\mathrm{CMB}}}(r,z) = \bar{\tau}_{\mathrm{eff}} \frac{v_{12}(r,z)}{c},
\label{eq:Tkszv12}
\eeq
where $\bar{\tau}_{\mathrm{eff}}$ is the average effective (i.e.,\ beam-convolved) optical depth of the cluster sample. 

In practice, measuring the pairwise kSZ signal consists of a two-step process: first, the CMB map is filtered to reduce the noise. This can be achieved by applying a \emph{matched filter} \citep{Haehnelt:1995dg}, which takes into account the spectral dependence of the noise as well as the spatial profile of the signal, and was applied in the analyses in \citet{Hand:2012ui,Soergel:2016mce}. Another filter to reduce noise is the \emph{compensated top-hat filter}, which simply computes the average signal within an aperture and subtracts the average signal in a ring with equal area around it \citep{Ade:2015lza,DeBernardis:2016pdv}. \citet{Flender:2015btu} show that the compensated top-hat filter can perform almost as good as the matched filter, depending on the aperture size $\theta$.

Second, the pairwise kSZ signal is extracted from the filtered map using the so-called \emph{pairwise estimator} which was originally introduced in \citet{Ferreira:1998id}, and re-written in \citet{Hand:2012ui} in terms of CMB temperature values,
\beq
\Delta T_{\mathrm{pkSZ}} (r,z) = -\frac{ \sum_{ij} c_{ij}\,T_{ij}}{\sum_{ij} c^2_{ij}},
\label{p_est}
\eeq
where the sum is taken over all pairs in the sample, $T_{ij}$ is the difference in filtered temperature values at the cluster locations, and $c_{ij}$ are geometric weights given by 
\beq
\label{weights}
c_{ij} \equiv {{\bf\hat r}_{ij}}\cdot\frac{{\bf\hat r}_i + {\bf\hat r}_j}{2} = \frac{(r_i - r_j)(1+\cos\theta)}{2\sqrt{r_i^2 + r_j^2 - 2r_ir_j\cos\theta}}.
\eeq 
Here, ${\bf\hat r}_i$ is the unit vector pointing to cluster $i$, ${\bf\hat r}_{ij}$ is the unit vector pointing from cluster $i$ to cluster $j$, $r_i$ is the comoving distance of cluster $i$, and $\theta$ is the angular separation between the two clusters. It can be mathematically shown that the pairwise estimator returns an unbiased estimate of the true pairwise velocity \citep{Ferreira:1998id}.

At large scale (i.e., in the linear regime), the pairwise velocity can be expressed as $v_{12}(r) = 2 \bar{b} \xi_{v\delta}(r)$ \citep{Keisler:2012eg}, where $\bar{b}$ is the mass-averaged halo bias (which can be measured via the cluster's auto-correlation), and $\xi_{v\delta}(r)$ is the matter-velocity correlation function, 
\beq
\xi_{v\delta}(r,z) = -a H f \int \dd k k P(k,z) j_1 (kr),
\label{eq:xiP}
\eeq
where $a$ is the scale factor, $H$ is the Hubble rate, $f$ is the growth function, and $j_1$ is the first spherical Bessel function. Thus, the large-scale pairwise kSZ measurement probes
\begin{equation}
\Delta T_{\mathrm{pkSZ}} \sim \bar{\tau}_{\mathrm{eff}} f \sigma_8^2,
\label{eq:ksz_tau_f}
\end{equation}
where $\sigma_8$ denotes the normalization of the matter power spectrum. From a cosmological perspective, the dependence on the growth function $f$ is arguably one of the most interesting features, because different models of gravity and dark energy predict a different $f(z)$. 

Note that the pairwise kSZ measurement based on galaxy surveys is affected by redshift-space distortions, which lead to small suppression of the signal at $\sim20-100\,\Mpc$ and a sign inversion at $\lesssim20\Mpc$, as seen in Fig.\,3 in \citet{DeBernardis:2016pdv}. This effect requires a more careful modeling, which has been discussed in e.g.,\ \citet{Okumura:2013zva,Sugiyama:2015dsa}.

\subsection{Cosmological implications}

The potential of kSZ measurements as a probe of cosmology and gravity has been discussed in e.g.,\ \citet{DeDeo:2005yr,HernandezMonteagudo:2005ys,Bhattacharya:2007sk,Kosowsky:2009nc,Keisler:2012eg,Ma:2013taq,Mueller:2014nsa,Mueller:2014dba,Alonso:2016jpy}. Here, we highlight a few illustrative examples from the recent literature.

\citet{Alonso:2016jpy} show that, with data from Stage\,IV experiments, it would be possible to measure the product of the Hubble rate and the growth rate, $f(z)H(z)$, to better than $1\%$ out to $z=1$ with the redshift bins of $\Delta z=0.1$. However, their conclusion hinges on the assumption that we will have precise knowledge of the optical depth, which, they argue, can be obtained using the scaling relation between the integrated tSZ $y$-parameter and $\tau$. \citet{Sugiyama:2016rue} argue that combining future kSZ and galaxy redshift survey data can reduce the marginalized $1\sigma$ errors on $f$, as well as on the Hubble rate, by $\sim$50-70\%, compared to the galaxy-only analysis. Given the degeneracy between $f$ and $\tau$ (as shown by their Fig.\,7), these constraints could be further improved by including an external prior on $\tau$. 

\citet{Mueller:2014nsa} perform a Fisher-matrix analysis to investigate how well future pairwise kSZ measurements can help constrain dynamical dark energy models and modified gravity. For the former, the dynamical dark energy is parametrized as the dark energy EOS, ${w=w_0 + (1-a)w_a}$ ($w_0=-1,\ w_a=0$ for the concordance $\Lambda$CDM model). For the latter, the modified gravity models are parametrized in terms of the growth function, $f(z)=\Omega_m(z)^{\gamma_{\mathrm{growth}}}$ with a free parameter $\gamma_{\mathrm{growth}}$ (general relativity predicts $\gamma_{\mathrm{growth}}\simeq0.55$). The authors find that combining data from a Stage\,III galaxy redshift survey (BOSS) with Stage\,III CMB data (Advanced ACTpol) will yield $1\sigma$ errors on $w_0$ and $w_a$ of $0.08$ and $0.26$, respectively, as well as 5\% constraints on $\gamma$. Although these authors assumed a prior of $\sim40\%$ on $\tau$, we highlight that these constraints could be significantly improved with a stronger prior on $\tau$. In particular, the authors find that a $\sim10\%$ prior on $\tau$ could enable Stage\,II and III CMB surveys to provide constraints that are competitive with respect to Stage\,IV constraints (see their Fig.\,8).

\citet{Mueller:2014dba} show that the scale-dependence of the pairwise kSZ signal can be used to constrain neutrino masses. In particular, the authors forecast 68\% upper limits on the sum of the neutrino masses, $\sum m_{\nu} = 290$ meV, $220$ meV, $96$ meV for Stage II, Stage III, and Stage IV surveys, respectively. The authors further show that percent-level constraints on $\tau$ will improve these constraints to 120 meV, 90 meV and 33 meV, respectively. For comparison, \citet{deHaan:2016qvy} find $\sum m_{\nu} = 140\pm80\,$meV, by combining data from SPT clusters with Planck CMB data and baryon acoustic oscillation data.

Furthermore, \citet{Keisler:2012eg} discuss the potential of the pairwise kSZ signal to constrain specific modified gravity models. In particular, the authors consider the normal-branch Dvali–Gabadadze–Porrati (DGP, \citealt{Dvali:2000hr}) model with preserving the expansion history to be that of the $\Lambda$CDM model \citep{Schmidt:2009sv}, which leads to a scale-independent modification of the growth function. The authors show that the pairwise velocity at linear scales can be up to 15\% higher in the DGP model, compared to the concordance $\Lambda$CDM model. The authors further demonstrate that the $f(R)$ model \citep{Carroll:2003wy}, which invokes a massive additional degree of freedom, generates a scale-dependent modification into the pairwise velocity, compared to the $\Lambda$CDM model.

In all of the cases, it is clear the science return from the pairwise kSZ signal depends strongly on our understanding of the optical depth. This stresses the motivation behind this work, which is to constrain the optical depth profile to $\lesssim 10\%$.  

\section{ICM Model}

The primary goal of this work is (1) to develop a physically motivated and computationally efficient semi-analytic model of the optical depth profiles and (2) to constrain the model of the optical depth profile using the state-of-the-art X-ray observations of galaxy groups and clusters. Specifically, we will adopt the semi-analytic ICM model described in \citet{Shaw:2010mn} (hereafter the \emph{Shaw model}), which is a modified version of the model by \citet{Ostriker:2005ff}, and extend it to model {\it gas cooling} to provide a better description of recent X-ray data, especially in cluster cores. 

\subsection{Dark matter halo structure}
The Shaw model assumes that the gas inside the cluster initially follows the dark matter density distribution, which is modeled as a Navarro-Frenk-White (NFW) profile \citep{Navarro:1996gj},
\beq
\label{eq:nfw}
\rho_{\mathrm{DM}}(r) =  \frac{\rho_s}{(r/r_s) (1+r/r_s)^{2}} \;\;,
\eeq
where $r_s$ is the NFW scale radius and $\rho_s$ is a normalization constant. The scale radius is related to the virial radius through the halo concentration $c$, defined as $c = R_{\mathrm{vir}/r_s}$, where the virial radius is the radius enclosing the virial mass,
\beq 
M_{\mathrm{vir}} = \frac{4}{3} \pi R_{\mathrm{vir}}^3 \Delta_{\mathrm{vir}} \rho_c(z) \;,
\eeq
where $\Delta_{\mathrm{vir}} = 18 \pi^2 + 82 (\Omega_M(z)-1) -39 (\Omega_M(z)-1)^2$ is the virial overdensity and $\rho_c(z)$ is the critical density at redshift $z$. We also use the overdensity mass $M_{\Delta}$, defined as
\beq
M_{\Delta} = \frac{4}{3} \pi R_{\Delta}^3 \Delta \rho_c(z),
\label{eq:Mdelta}
\eeq
where $R_{\Delta}$ is the radius within which the enclosed mean density is $\Delta$ times the critical density of the universe. Current X-ray observations measure the gas density and temperature profiles roughly out to $R_{500}$. We define $\theta_{\Delta}$ as the angle subtended by $R_{\Delta}$ on the sky, i.e.\ $\theta_{\Delta} = R_{\Delta}/d_A(z)$, where $d_A(z)$ is the angular diameter distance. In this work, we assume that each halo has a concentration that is determined by its mass and redshift, following the relation by \citet{Duffy:2007mb},
\beq 
c(M_{\mathrm{vir}},z) = 7.85 \left(\frac{M_{\mathrm{vir}}}{2\times10^{12} \msol}\right)^{-0.081} (1+z)^{-0.71}\;.  
\eeq 

\subsection{Star formation}

\begin{table}
\begin{center}
\begin{tabular}{|c|c|c|}
\hline
Reference & $10^{2}f_*$ & $S_*$ \\ \hline \hline
\citet{0004-637X-591-2-749} & $1.64^{+0.10}_{-0.09}$& $0.26\pm 0.09$ \\
\citet{Gonzalez:2007aj} & $2.02\pm0.37$ & $0.64 \pm 0.13$ \\
\citet{Giodini:2009qf} & $2.58 \pm 0.05$ & $0.37\pm 0.04$ \\
\citet{Leauthaud:2011gw}$^{\dagger}$ & 1.2 - 2.5 at $M_{500}=10^{13}\Msol$ & - \\
& 0.57-1.5 at $M_{500}=10^{14}\Msol$ & -  \\
\citet{Budzynski:2013obt} & $0.912\pm0.06$ & $0.11 \pm 0.14$ \\
\hline
\end{tabular}
\end{center}
\caption{\label{table_fstar} Different values for $(f_*,S_*)$ reported in the literature. $^{\dagger}$\citet{Leauthaud:2011gw} do not report an estimate of $S_*$, but their reported values for $f_*$ at different masses are consistent with $S_* = 0$.}
\label{tab:fstar}
\end{table}

We assume that some fraction of the gas inside the halo has radiatively cooled and formed stars. Specifically, we assume that the stellar fraction $F_* = M_*/M_{500}$, i.e., the ratio of stellar mass to halo mass within $R_{500}$, depends only on the mass of the halo, and follows a power-law:
\beq
F_*(M_{500}) = f_*\left(\frac{M_{500}}{3\times10^{14}\Msol}\right)^{-S_*}.
\eeq
Our stellar model has thus 2 free parameters that control the normalization and the slope of the $F_*-M$ relation. Various different values for these parameters are reported in the literature, and we summarize some of them in Table\,\ref{table_fstar}. The values reported for $f_*$ vary from just below 1\% \citep{Budzynski:2013obt} to 2.58\% \citep{Giodini:2009qf}. The values for the slope $S_*$ vary from 0.64 \citep{Gonzalez:2007aj} to a value consistent with $0$ \citep{Leauthaud:2011gw}. Given the vastly different values found in the literature, here we do not implement any particular stellar model, but instead marginalize over $f_*$ and $S_*$, taking the literature values in order to inform our priors. In particular, we choose a flat prior $0\le S_*\le0.64$ and $0.01\le f_*\le0.03$, bracketing the values reported the literature.

\subsection{Gas distribution}
We assume that the gas inside the dark matter halo rearranges itself into a state of hydrostatic equilibrium (HSE), which is described by the differential equation
\beq
\frac{dP_{\mathrm{tot}}(r)}{dr} = -\rho_g(r) \frac{d\Phi(r)}{dr} \;,
\eeq
where $P_{\mathrm{tot}}$ is the total (thermal + non-thermal) pressure, $\rho_g$ is the gas density, and $\Phi$ is the dark matter NFW potential. The solution to this equation can be written as
\begin{eqnarray}
P_{\mathrm{tot}}(r) &=& P_0 \theta(r)^{n+1}  \\
\rho_{g}(r) &=& \rho_0 \theta(r)^{n} \;,
\end{eqnarray}
where $\theta(r)$ is the polytropic variable
\beq
\theta(r) = 1 + \frac{\Gamma-1}{\Gamma}\frac{\rho_0}{P_0}(\Phi_0 - \Phi(r)) \;,
\eeq
and $\Phi_0$ is the central potential of the cluster, ${\Gamma=1+1/n}$ is the \emph{adiabatic index}, and $n$ the \emph{polytropic index}. For instance, an isothermal fluid has $\Gamma=1$, a non-relativistic isentropic fluid has $\Gamma=5/3$, and a relativistic isentropic fluid has $\Gamma=4/3$. Hydrodynamical simulations suggest that the ICM follows $\Gamma\approx 1.2$ \citep{Ostriker:2005ff,Shaw:2010mn,Battaglia:2012cq} --- except in the cool core region as explained below. Note that, while our model includes the non-thermal pressure as outlined in \S2.2.5 of \citet{Shaw:2010mn}, the non-thermal pressure only affects the thermal temperature structure and has no impact on the optical depth profile. 

The normalization of the model (i.e., $P_0$ and $\rho_0$) is determined through the energy constraint equation, 
\beq
E_{g,f} = E_{g,i} + \epsdm |E_{\mathrm{DM}}| + \epsfb M_* c^2 + \Delta E_p \;,
\eeq
where the left hand side of the equation is the final energy in the ICM, and the right hand side consists of the following terms:
\bim
\item $E_{g,i}$ is the initial total energy in the ICM, which is simply the sum of the kinetic and potential energy of the dark matter halo, scaled by the cosmic baryon fraction (see Eq.\,10 in \citet{Shaw:2010mn}).
\item $\epsdm |E_{\mathrm{DM}}|$ is the energy introduced into the ICM during major halo mergers via dynamical friction heating, where $E_{\mathrm{DM}}$ is the total energy in the dark matter halo (i.e., the sum of kinetic and potential energy), and the parameter $\epsdm$ describes how much of that energy is induced into the ICM. While \citet{Bode:2009gv} suggest a value of $\epsdm=0.05$, based on the hydrodynamical simulations of \citet{McCarthy:2007ia}, the exact value of $\epsdm$ remains uncertain and likely depends on other factors such as the environment and merger history of a given halo. Here, we leave $\epsdm$ as a free parameter in the likelihood analysis.
\item $\epsfb M_* c^2$ is the energy injected into the ICM due to feedback from SNe and AGN, where $M_*$ is the stellar mass and $\epsfb$ is a free parameter that describes how much of the energy in stars is transformed into feedback energy. 
\item $\Delta E_p$ is the work done by the gas as it expands relative to its initial state. 
\eim

\subsection{Gas Cooling}
Hydrodynamical simulations show that an adiabatic index of $\Gamma \approx 1.2$ provides a good description of the ICM over a large range of scales, except in the high-density cluster core in which gas cooling causes the adiabatic index to become much lower (Lau \& Nagai, in prep). \citet{McDonald:2013fka} find a strong redshift evolution in the gas density profile, where the normalized central density increases by an order of magnitude from $z\sim1$ to $z\sim0$. Cooling was not modeled in \citet{Shaw:2010mn}, but is modeled in our new model, and it is particularly important for modeling the ICM in the central regions of galaxy groups and clusters. 

In order to account for cooling, we introduce 3 more degrees of freedom: a breaking point $\xbr=r_{\rm break}/R_{500}$, which controls at which point the adiabatic index breaks, and a new adiabatic index $\Gamma^{\prime}$ inside the broken region, $x<\xbr$. In order to take into account the redshift-dependence, we model the adiabatic index inside the broken region as
\beq
\Gamma^{\prime}(z) = \tilde{\Gamma}(1+z)^{\gamma}, 
\eeq
where $\gamma$ controls how strongly cooling effects evolve with redshift. For positive $\gamma$, the adiabatic index inside the core region, $\Gamma^{\prime}(z)$, decreases with redshift, and vice versa. 

\section{The $\tau-M$-relation}
\label{sec:tmr}

\begin{figure*}
\centering
\includegraphics[scale=0.7]{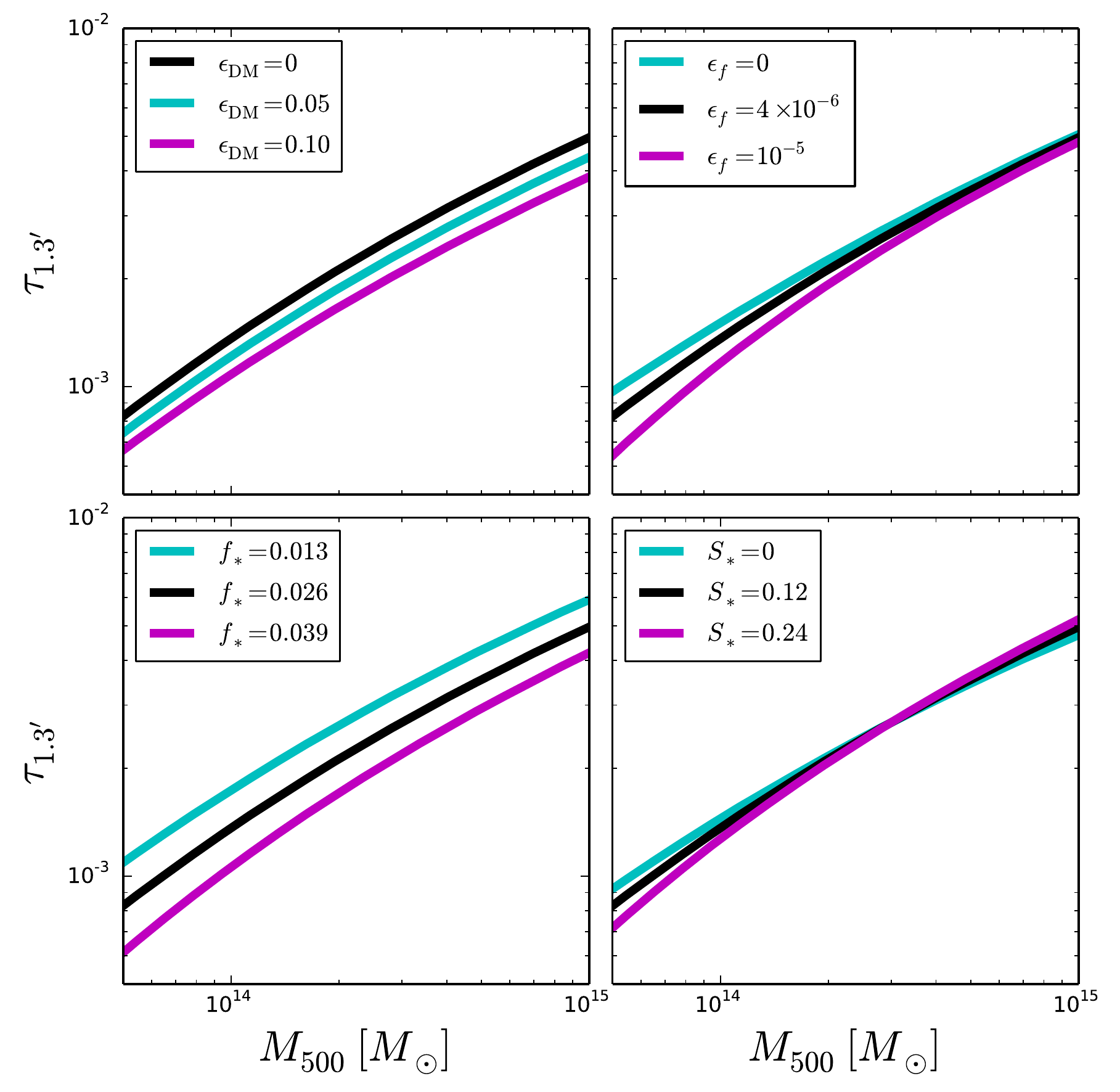}
\caption{\label{fig:tau-M_model_dependence} Dependence of the $\tau-M$ relation on the different parameters in our model, dark matter feedback (top left), feedback from AGN/SNe (top right), stellar fraction (bottom left), and the slope in the relation between stellar fraction and halo mass (bottom right). In each panel, the black line shows our best-fit model, as determined via MCMC analysis described in Section\,\ref{sec:obs_cali}. We find that the normalization of the $\tau-M$ relation depends strongly on feedback parameters and the stellar fraction, while the slope of the $\tau-M$ relation depends on the energy feedback from AGN/SNe and the slope in the stellar model.}
\end{figure*}

\begin{figure}
\centering
\includegraphics[scale=0.7]{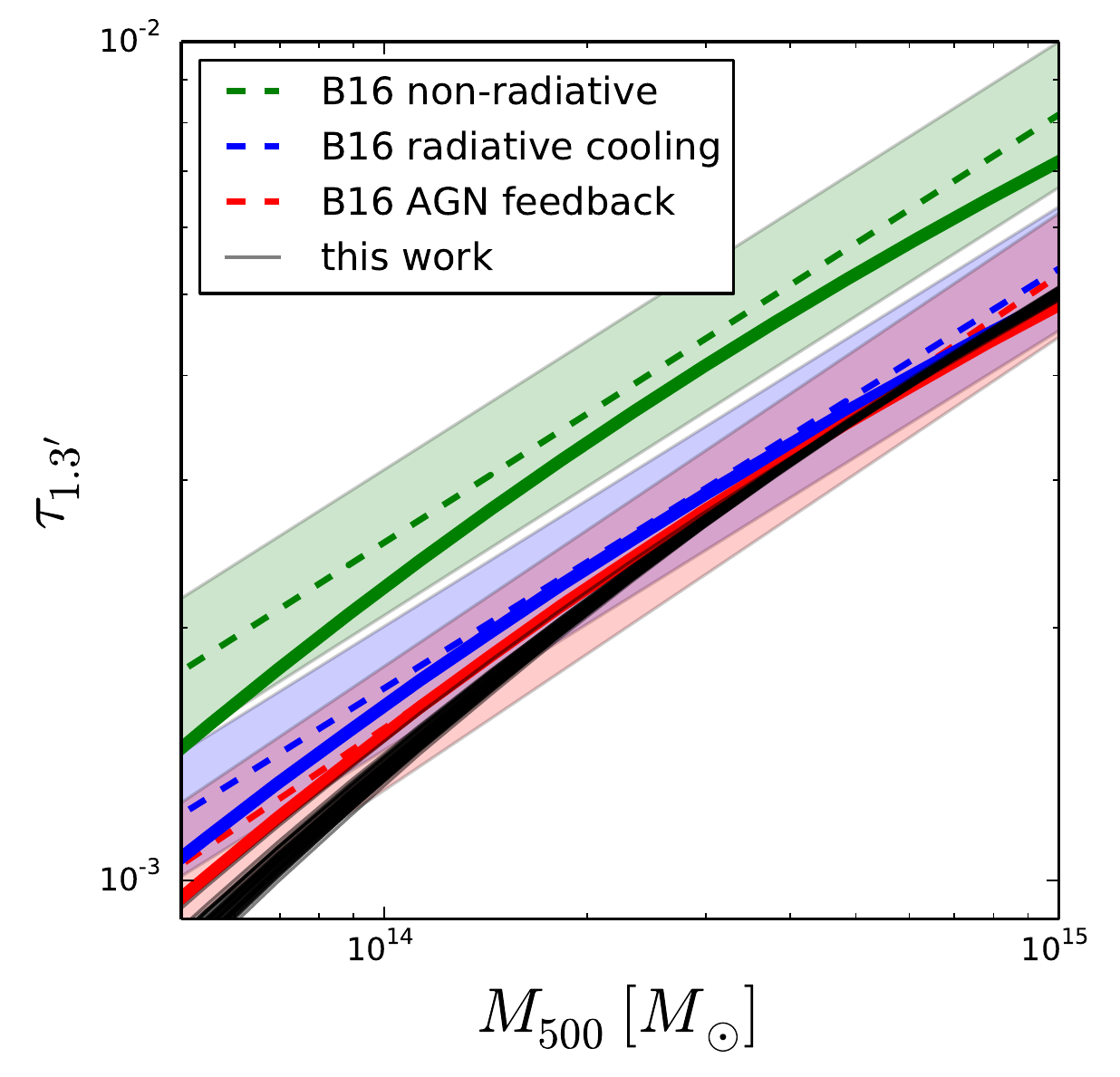}
\caption{\label{fig:comp_to_Battaglia} Reconstruction of the results from B16. The dashed lines and shaded regions show the results for the $\tau-M$ relation from B16, while the solid lines show the reconstruction using our model with the appropriate parameters (see text for details). The black lines are 50 models with parameters from our MCMC chain, indicating the range of our observationally calibrated model.}
\end{figure}

Recently, B16 used the output from hydrodynamical simulations to show that there are simple scaling relations that could be used to break the degeneracy between optical depth and cosmology constraints. Specifically, B16 found that there exists a power-law scaling relation between $\tau_{\theta}$ measured within an aperture of angular size $\theta$ (defined in Eq.\,\ref{eq:tautheta}), and the Compton-$y$ parameter measured within the same aperture, $y_{\theta}$, which could be measured via the cluster's tSZ signature. Another scaling relation exists between $\tau_{\theta}$ and the halo mass $M_{500}$. B16 studied the normalization of these scaling relations for three different sets of cluster physics: one model without any cooling and star-formation (`non-radiative'), one model with cooling and feedback from SNe (`radiative cooling'), and one model with additional feedback from AGN (`AGN feedback').

The advantage of our semi-analytic approach is that we can explore a large number of ICM models with considerably less computational cost, compared to hydrodynamical simulations. In Fig.\,\ref{fig:tau-M_model_dependence} we demonstrate the dependence of the $\tau-M$ on the different parameters in our model. We find that the normalization of the $\tau-M$ relation depends strongly on the amount of dynamical friction heating from dark matter feedback, as well as the stellar fraction (see left two panels in Fig.\,\ref{fig:tau-M_model_dependence}). The AGN/SNe feedback parameter changes both the normalization of the $\tau-M$ relation as well as its slope. This finding is consistent with B16, who also find a steeper slope in their AGN run (0.54) compared to their non-radiative run ($0.5$). The slope further depends on the slope in the $f_*-M$-relation, as seen in the bottom right panel of Fig.\,\ref{fig:tau-M_model_dependence}. A steeper slope in the $f_*-M$ relation creates a steeper slope in the $\tau-M$ relation. 

In Fig.\,\ref{fig:comp_to_Battaglia} we demonstrate that we are able to reproduce the results of the $\tau-M$ relations presented in B16 by appropriate adjustment of parameters in our model. We find good agreement between the B16 `non-radiative' run and our model with $f_* = \epsfb = \epsdm =0$. In order to reconstruct the B16 `radiative cooling' run, we set $f_* = 0.024$ and $S_*=0$, in order to match the stellar model from B16. Finally, in order to reconstruct the `AGN feedback' run, we set $\epsfb=4\times 10^{-6}$.

The set of black lines in Fig.\,\ref{fig:comp_to_Battaglia} shows the range of our observationally calibrated model (described below), which is in broad agreement with the B16 `AGN feedback' run, however with a slightly steeper slope due to our steeper slope in the $f_*-M$ relation. These results demonstrate flexibility and capability of reproducing the results of modern hydrodynamical simulation with varying input cluster astrophysics. 

\section{Observational calibration of the model}
\label{sec:obs_cali}

\begin{figure*}
\centering
\includegraphics[scale=0.8]{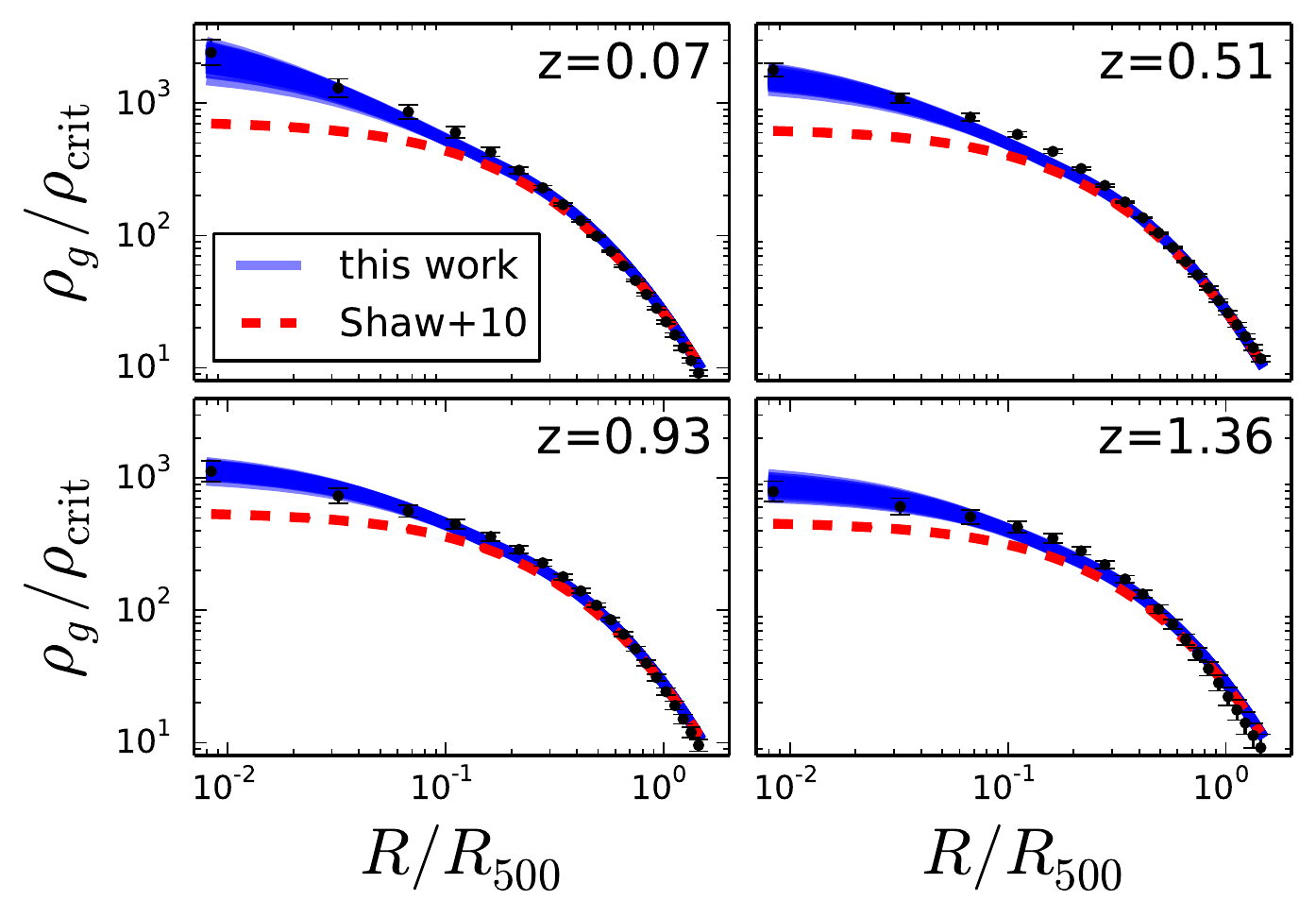}
\caption{\label{fig:density_profiles} Gas density profiles from \citet{McDonald:2013fka}, along with 50 lines from our model, using parameters from the MCMC chain, showing the range of the model (blue lines). Also shown is the fiducial model from \citet{Shaw:2010mn} (red dashed line), which under-predicts the central gas density because of the lack of a cooling mechanism.}
\end{figure*}

\begin{figure}
\centering
\includegraphics[scale=0.7]{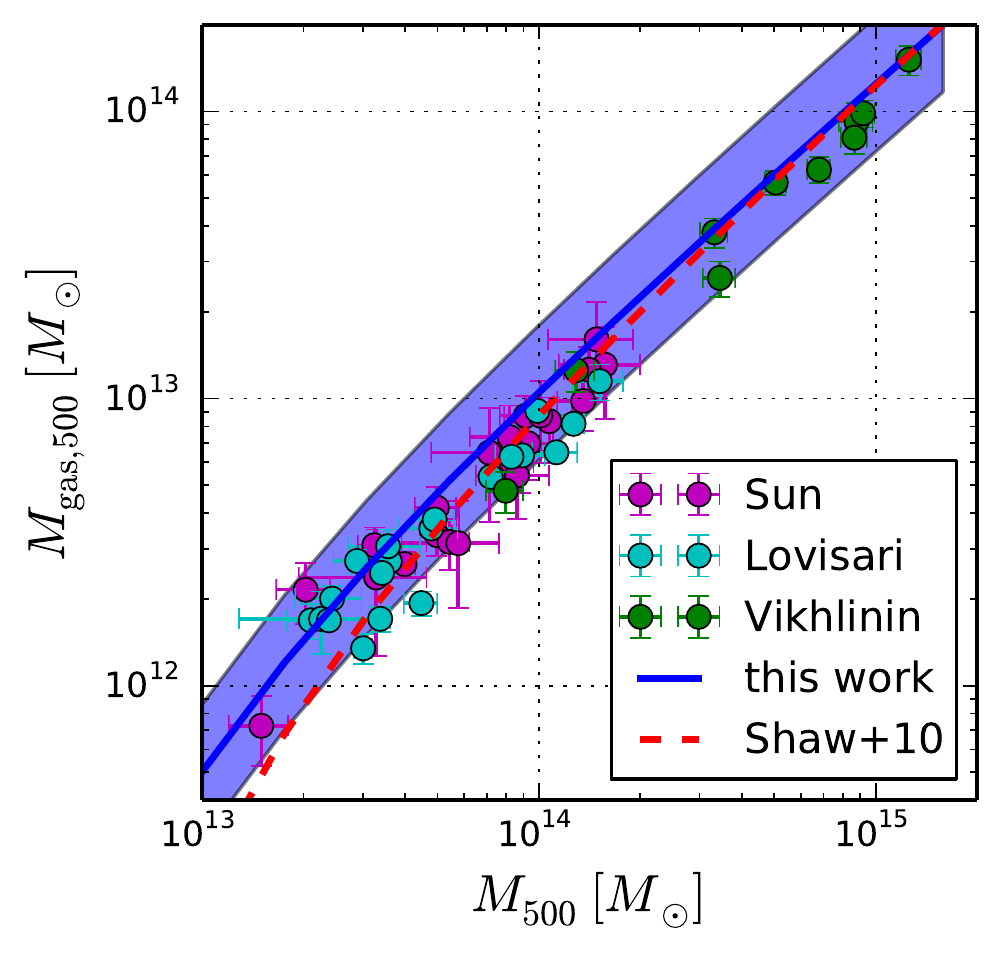}
\caption{\label{fig:mgas} $M_{\mathrm{gas}}-M$ relations from \citet{0004-637X-640-2-691}, \citet{Sun:2008eh} and \citet{2015A&A...573A.118L}. The blue line shows our best-fit model, and the shaded region shows the $2\sigma_{\ln M_{\mathrm{gas}}}$ log-scatter. Given the measurement errorbars, the model provides a good description of the data. For comparison, we also show the result from the original \citet{Shaw:2010mn} model (red dashed line), which has a slightly stronger steepening toward the low-mass end due to the higher value of $S_*$ adopted in that model. }
\end{figure}

\begin{figure*}
\centering
\includegraphics[scale=1.0]{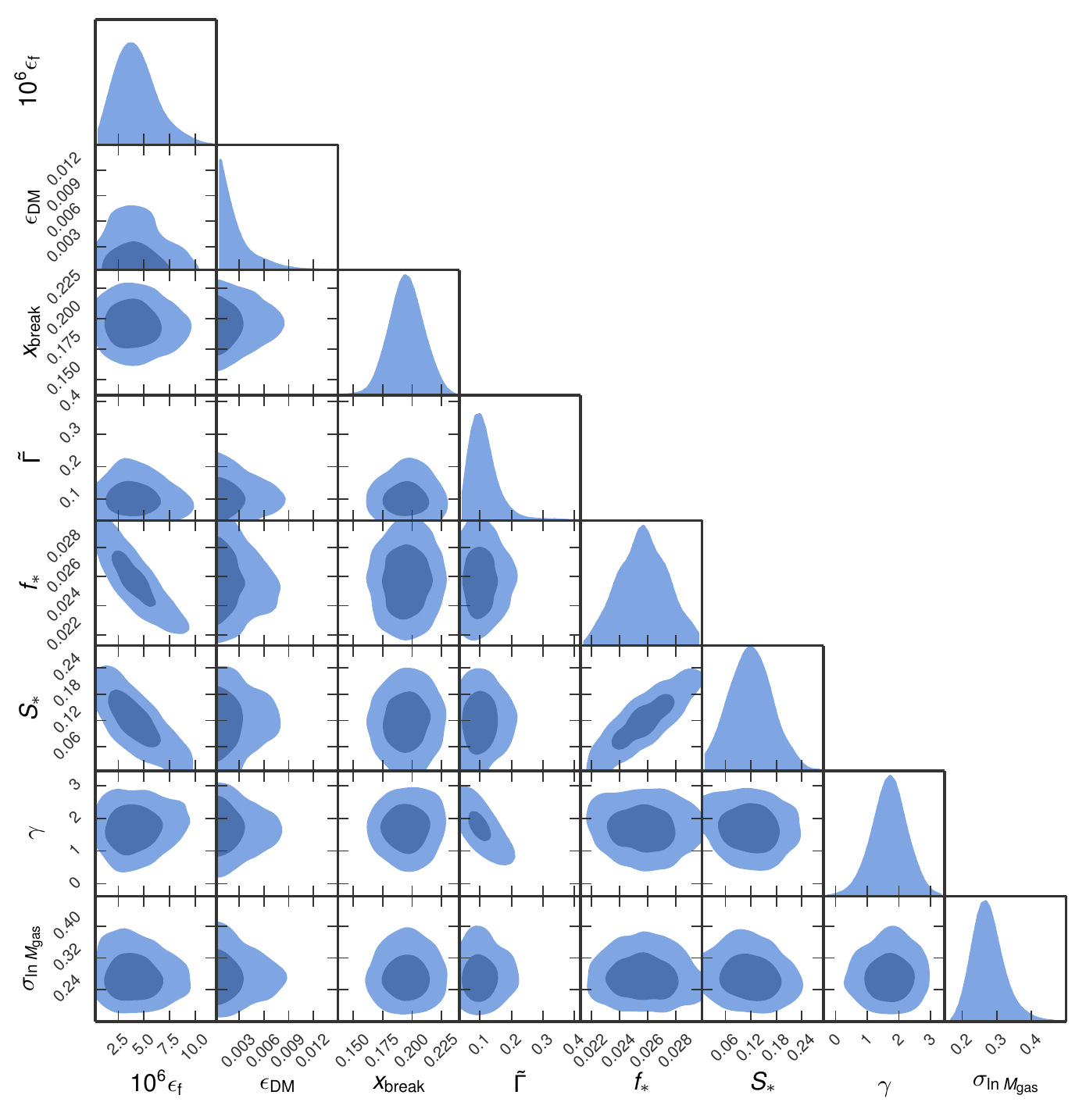}
\caption{\label{fig:contours} Contour plot from our MCMC likelihood analysis. The data used in this analysis are gas density profiles from \citet{McDonald:2013fka} and measurements of the $M_{\mathrm{gas}}-M$ scaling relation from \citet{0004-637X-640-2-691}, \citet{Sun:2008eh} and \citet{2015A&A...573A.118L}. The data has enough constraining power to limit the parameter space for most parameters in our model, however it has some degeneracies between $f_*$, $S_*$, and $\epsfb$, which could be broken with additional measurements of the stellar content of groups and clusters. We report the best-fit values in Table\,\ref{table_values_MCMC}.}
\end{figure*}

\subsection{Observational data sets}
\label{sec:obsdata}
In order to calibrate our model we implement two types of data into our likelihood analysis. We use recent X-ray measurements of galaxy groups and clusters, including gas density profiles of massive clusters by \citet{McDonald:2013fka} as well as $\mgasfh$ measurements of groups and clusters from \citet{0004-637X-640-2-691}, \citet{Sun:2008eh} and \citet{2015A&A...573A.118L}:

\paragraph{\citet{McDonald:2013fka}} presented the results of an X-ray analysis of 83 galaxy clusters that were selected in the 2500\,sq.deg. SPT survey and observed with the Chandra X-ray Observatory. The authors measured the average shape of the gas density profile over the radial range $0<r/R_{500}<1.5$. We combine this data set with 8 additional clusters at $z>1.2$ that were obtained as a separate Chandra program (PI: McDonald) and will be presented in an upcoming paper (McDonald et al, in prep). The combined data set is binned into four different cluster-subsamples at redshifts 0.07, 0.51, 0.93, and 1.36, with average masses $M_{500}/(10^{14}\Msol)$ of 5.76, 5.09, 4.17, and 2.84, respectively.

\paragraph{\citet{0004-637X-640-2-691}} presented gas and total mass profiles for 13 relaxed clusters at low redshift ($0.0162<z<0.2302$), spanning a temperature range of 0.7–9\,keV, derived from Chandra data. For 10 of those clusters the authors report measurements of $\fgasfh$ and $M_{500}$, which we use to derive the $M_{\mathrm{gas},500}-M_{500}$ relation in that sample. The mass range of that sample is $0.77 < M_{500}/10^{14}\Msol < 10.74$.

\paragraph{\citet{Sun:2008eh}} presented a systematic analysis of 43 nearby galaxy groups in the mass range $10^{13}<M_{500}/\Msol<10^{14}$ and redshift range $0.012 < z < 0.12$, based on Chandra archival data. These data contain measurements of $\fgasfh$ for 23 objects. By combining these measurements with their hydrostatic mass estimates for $M_{500}$, we obtain the $M_{\mathrm{gas},500}-M_{500}$ relation for this sample, which we use in our likelihood analysis.

\paragraph{\citet{2015A&A...573A.118L}} analyzed XMM-Newton observations for a complete sample of galaxy groups selected from the ROSAT All-Sky Survey. The data consists of 20 objects in the redshift range $0.012<z<0.034$ and the mass range $2.07\times 10^{13}<M_{500}<1.44\times 10^{14}$. For all of these objects the authors report measurements of $M_{\mathrm{gas},500}$ and $M_{500}$, which we use in our likelihood analysis.

\subsection{Correcting for the hydrostatic mass bias}
All of the data used in our likelihood analysis contain estimates of $M_{500}$ that were derived assuming that the ICM is in HSE with the gravitational potential of groups and clusters. However, this is only true for the thermal component of the gas, not the non-thermal component due to internal motions in galaxy clusters. This leads to a bias in the hydrostatic mass estimate, known as HSE bias. To model the HSE mass bias, we use the semi-analytic model of the non-thermal pressure, which has been calibrated to hydrodynamical simulations \citep{Shi:2016}. This model predicts the HSE mass bias of 1\%, 4\%, and 14\% for groups and clusters with $M_{500}=10^{13}$, $10^{14}$, and $10^{15}\Msol$, respectively. We use this model to remove the HSE bias from the mass estimates used in our likelihood analysis.

\subsection{MCMC analysis}
We constrain the parameter space of the model, in light of the data, using the Markov-Chain Monte Carlo (MCMC) algorithm, with the likelihood $\mathcal{L}=\exp(-\chi^2/2)$. We compute the total $\chi^2$ as the sum of the $\chi^2$ contributions from the 4 different data sets described in \S\ref{sec:obsdata}. For the McDonald data, we approximate the $\chi^2$ as the sum of deviations of the model from the data over all radial bins (i.e., we neglect the cross-correlation between radial bins):
\beq
\chi^2 = \sum_i \frac{ (\rho_{i,\mathrm{data}} - \rho_{i,\mathrm{model}} )^2 } {\sigma^2_{i,\mathrm{data}}}.
\eeq
For the $\mgas-M$ relations we compute the $\chi^2$ as 
\beq
\chi^2 = \int \dd \mgas p_{\mathrm{data}}(\mgas,\sigma_{\mgas}) p_{\mathrm{model}} (\mgas, \sigma_{\ln M_{\mathrm{gas}}}).
\eeq
Here, we assume that the probability distribution function in the data, $p_{\mathrm{data}}$, is a Gaussian function with mean $\mgas$ and standard deviation $\sigma_{\mgas}$, which is the error reported in the data. In addition, we assume that the $M_{\mathrm{gas}}-M$ relations have an intrinsic log-scatter with a width that is specified by the parameter $\sigma_{\ln M_{\mathrm{gas}}}$, i.e.,\ $p_{\mathrm{model}}$ is a log-normal distribution. The physical reason for introducing the parameter $\sigma_{\ln M_{\mathrm{gas}}}$ is that not all clusters with a given mass and redshift are expected to contain exactly the same gas mass, but instead there exists an object-to-object scatter that is sourced by diversity in environment and formation history of groups and clusters. Here, we leave $\sigma_{\ln M_{\mathrm{gas}}}$ as a free parameter and marginalize over it. 

Our final model has thus 8 free parameters: $\epsfb$, $\epsdm$, $\xbr$, $\tilde{\Gamma}$, $f_*$, $S_*$, $\gamma$, and $\sigma_{\ln M_{\mathrm{gas}}}$. We will present the results of the MCMC analysis in the next subsection.

\subsection{Constraints on the ICM model}

Fig.\,\ref{fig:density_profiles} shows that our new model (blue line, which makes an attempt to model the effect of gas cooling through modification of the effective EOS in high-density cores) provides a much better description of the observed gas density profiles, compared to the original \citet{Shaw:2010mn} model (red dashed line, which does not make an attempt to model the effects of gas cooling). 

Fig.\,\ref{fig:mgas} further shows the $M_{\mathrm{gas}}-M$ relations from the 3 different data sets \citep{0004-637X-640-2-691,Sun:2008eh,2015A&A...573A.118L}, which cover a wide mass range ($10^{13}<M_{500}/\Msol<10^{15}$). The blue line shows our best-fit model, and the shaded region shows the $2\sigma_{\ln M_{\mathrm{gas}}}$ log-scatter. For comparison, we also show the original \citet{Shaw:2010mn} model (red dashed line). The two models are very similar at the high-mass end, but diverge at the low-mass end, where our new model predicts a slightly larger gas mass for any given halo mass $M_{500}$. This difference originates primarily from the different slope in the $f_*-M$ relation; namely, \citet{Shaw:2010mn} adopted a much steeper slope ($S_*=0.37$), compared to our best-fit slope ($S_*=0.12$), and therefore predict a higher stellar fraction (and thus lower gas mass) at low-mass groups, compared to our model.

Fig.\,\ref{fig:contours} presents the constraints from the data on the 8-dimensional parameter space of our model. The best-fit values and 95\% confidence intervals are given in Table\,\ref{table_values_MCMC}. Based on these results, we draw the following conclusions about the ICM parameters of our model: 

\begin{table}
{\footnotesize
\begin{center}
\begin{tabular}{|c|c|}
\hline
parameter & value \\ \hline \hline
$10^{6}\epsfb$ & $3.97^{+4.82}_{-2.88}$\\
$\epsdm$ & $<0.0064$\\
$x_{\mathrm{break}}$ & $0.195^{+0.025}_{-0.024}$\\
$\tilde{\Gamma}$& $0.10^{+0.11}_{-0.05}$\\
$f_*$ &$0.026\pm0.003$\\
$S_*$ &$0.12\pm0.1$\\
$\gamma$ &$1.72^{+0.95}_{-1.04}$\\
$\sigma_{\ln M_{\mathrm{gas}}}$ &$0.27^{+0.11}_{-0.07}$\\ \hline
\end{tabular}
\end{center}
}
\caption{\label{table_values_MCMC} Best-fit parameters with 95\% confidence intervals of our 8-parameter model.}
\end{table}

\paragraph{AGN/SNe feedback} The data prefers a non-zero amount of feedback from AGN and SNe,  $10^{6}\epsfb=3.97^{+4.82}_{-2.88}$ (95\% confidence level). This value is larger than the value of $1$ adopted in the fiducial model in \citet{Shaw:2010mn} (although consistent within the 95\% confidence interval), but smaller than the value of $39$ adopted in \citet{Ostriker:2005ff} and $30-50$ suggested in \citet{Bode:2009gv}. As shown in Fig.\,\ref{fig:contours}, the amount of AGN/SNe feedback is strongly degenerate with the stellar fraction $f_*$, and with the slope in the stellar model, $S_*$. This degeneracy is expected because our model is only sensitive to the total feedback energy, $\epsfb M_*$, and could be broken with additional measurements of the $f_*-M$ relation, derived from the measurements of the stellar content of groups and clusters \citep{Bode:2009gv}.

\paragraph{Dark matter feedback} The data is consistent with zero feedback from dynamical friction heating from major halo mergers. We find that $\epsdm<0.0064$ with a confidence level of 95\%. This is much lower than the fiducial value of 0.05 assumed in \citet{Bode:2009gv} and \citet{Shaw:2010mn}. Ultimately, this number could be better constrained using hydrodynamical simulations.

\paragraph{Gas cooling} The data prefers a broken adiabatic index that breaks at $x=0.2$ from the standard value $\Gamma=1.2$ to a much smaller value $\Gamma^{\prime}=0.1\times(1+z)^{\gamma}$ with a strong redshift evolution with $\gamma=1.72^{+0.95}_{-1.04}$. This means that a cluster at $z=1$ has $\Gamma^{\prime}=0.33$, and a cluster at $z=0$ has $\Gamma^{\prime}=0.1$. This strong redshift evolution of the central density profile has been pointed out previously in \citet{McDonald:2013fka}. Our semi-analytical model provides a way to quantify that evolution of cool cores through the time-dependent change in the effective EOS, defined using the $\gamma$ parameter. Upcoming X-ray measurements of the gas density profile should help improve the constraints on the evolution in the effective EOS of cluster cores.

\paragraph{Stellar fraction} Our fits to the X-ray gas density and $\mgasfh$ data prefer a stellar fraction of $f_*=0.026 \pm 0.003$, and a slope of $S_*=0.12 \pm 0.1$. Our best-fit value for $f_*$ is consistent with the results by \citet{Giodini:2009qf}, who find $f_*=0.0258\pm0.0005$, but our analysis prefers a shallower slope than $S_*=0.37 \pm 0.04$ reported by \citet{Giodini:2009qf}. Our results are thus more in line with a recent analysis by \citet{Leauthaud:2011gw}, who find that the slope is much shallower than the slope reported in \citet{Giodini:2009qf}. Improved measurements of the $f_*-M$ relation will be important for breaking the degeneracy between $f_*$ and $S_*$ in our model. 

\paragraph{Scatter in $M_{\mathrm{gas}}-M$} We find that the data prefers a non-zero log-scatter of 0.27 in the $M_{\mathrm{gas}}-M$ scaling relation. This number quantifies the object-to-object scatter due to the fact that different clusters have different formation histories and live in different environments. 
Note that part of the cluster sample in this analysis is biased: \citet{0004-637X-640-2-691} only study relaxed massive clusters, while \citet{Sun:2008eh} and \citet{2015A&A...573A.118L} only study low-redshift systems. Therefore, our estimate of $\sigma_{\ln M_{\mathrm{gas}}}$ is possibly biased low, which could lead to an additional uncertainty in the estimate of the optical depth of individual objects.

\subsection{Constraints on the $\tau$ profile and integrated $\tau$}
\label{sec:constraints_on_tau_profile}

\begin{table}
{\footnotesize
\begin{center}
\begin{tabular}{|c|c|c|c|}
\hline
$M_{500}/\Msol$ & $10^3\tau(r=0)$ & $10^3\tau(r=R_{500})$ & $10^3\tau_{500}$ \\ \hline \hline
$5\times10^{13}$ & $2.84\pm0.34$ (12\%) & $0.37\pm0.02$ (4.7\%) & $0.71\pm0.04$ (6.3\%) \\
$10^{14}$ & $3.97\pm0.29$ (7.2\%) & $0.28\pm0.02$ (3.5\%) & $0.96\pm0.02$ (3.9\%)\\
$5\times10^{14}$ & $7.27\pm0.27$ (3.6\%) & $0.90\pm0.01$ (1.2\%) &$1.82\pm0.01$ (0.8\%)\\
$10^{15}$ & $9.32\pm0.37$ (4\%) & $1.18\pm0.01$ (0.8\%) & $2.36\pm0.02$ (1\%) \\\hline
\end{tabular}
\end{center}
}
\caption{\label{table_tau_values} Constraints on $\tau(r=0)$, $\tau(r=R_{500})$, and $\tau_{500}$, for different masses, all at $z=0.5$. The errorbars quoted are the 95\% confidence intervals.}
\end{table}

\begin{figure*}
\centering
\includegraphics[scale=0.8]{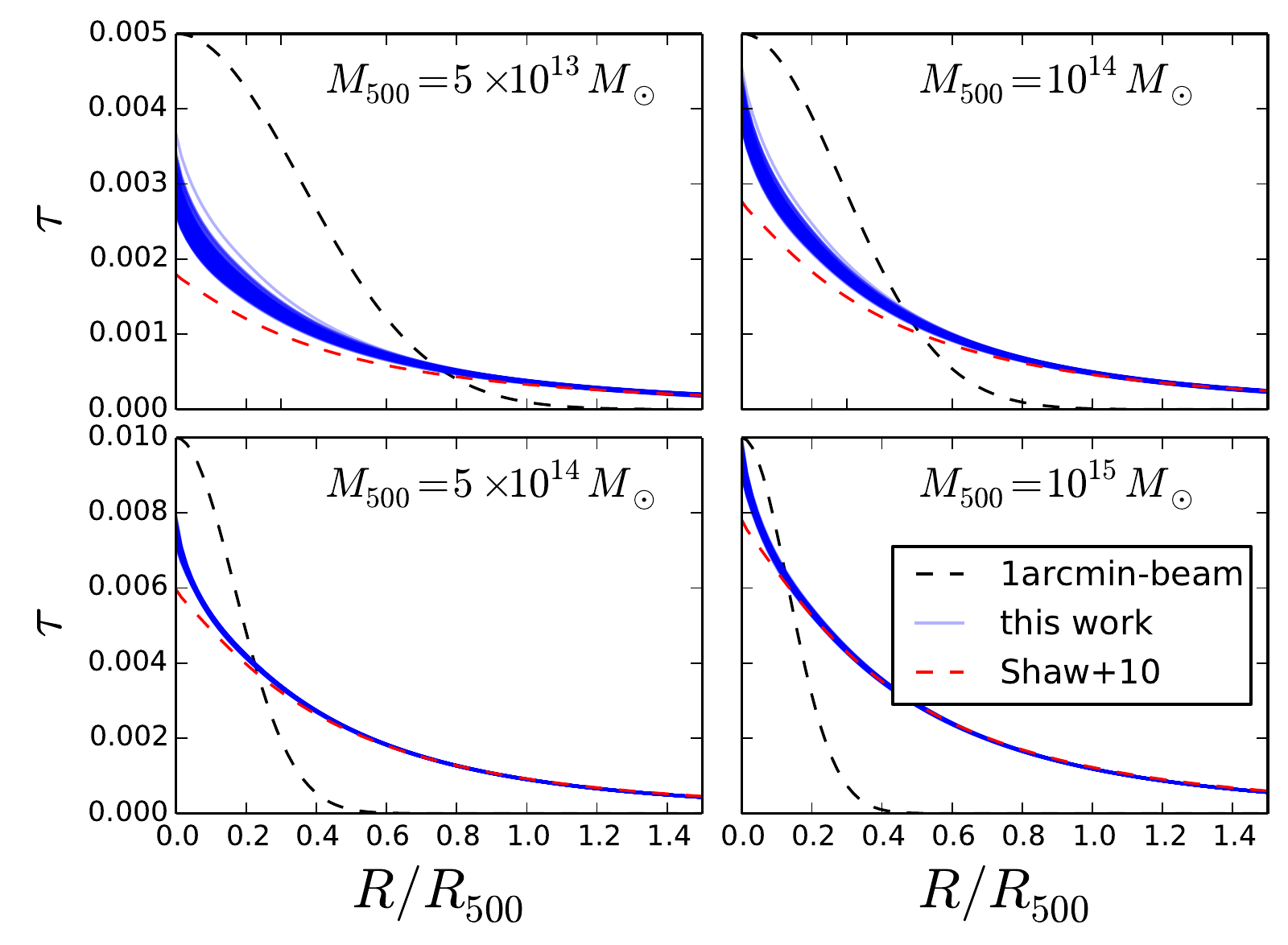}
\caption{\label{fig:tau_prof} Optical depth profiles for groups and clusters in the mass range $5\times 10^{13}-10^{15}\Msol$, all at $z=0.5$. The blue lines are 50 lines with parameters from our MCMC chains, illustrating the modeling uncertainty. The modeling uncertainty is higher at lower masses and smaller radii, such that the highest modeling uncertainty is at the central optical depth at the $5\times10^{13}\Msol$ object, which we determine to be 12\% at the 95\% confidence level. For comparison, we also show the fiducial model from \citet{Shaw:2010mn}, which predicts a lower central optical depth due to the lack of a cooling mechanism. The differences between our model and the 2010 model are however small on scales larger than the 1\,arcmin CMB instrument beam (black dashed line).}
\end{figure*}

\begin{figure*}
\centering
\includegraphics[scale=0.8]{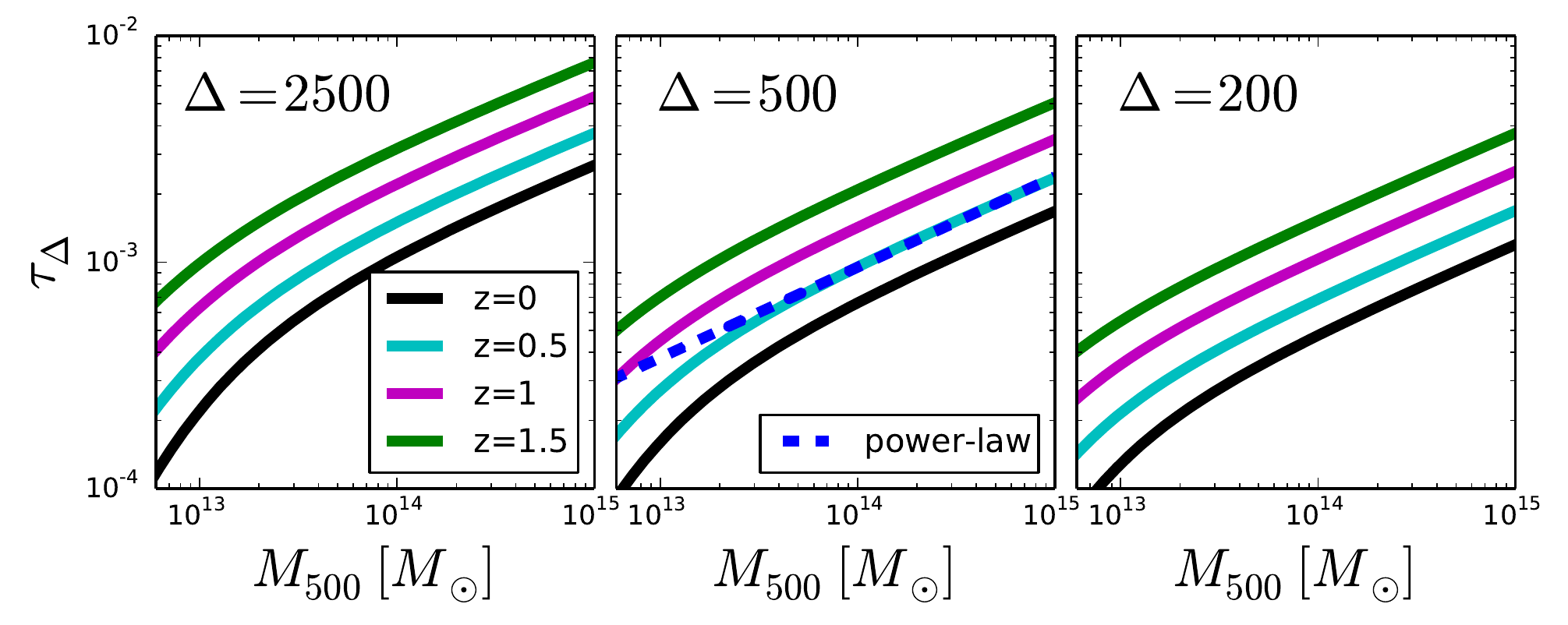}
\caption{\label{fig:tau_Delta_z} $\tau_{2500}$, $\tau_{500}$, and $\tau_{200}$ as a function of $M_{500}$ for various redshifts. The average $\tau_{\Delta}$ within an aperture $\theta_{\Delta}$ increases with increasing $\Delta$, i.e.\ decreasing aperture size, because in that case we probe more of the central region of the profile. The plot also shows that, at fixed mass and $\Delta$, clusters at higher redshift have a higher $\tau$. The blue dashed line in the middle panel shows the power-law approximation $\tau_{500} \simeq 0.95\times10^{-3}\,[M_{500}/(10^{14}\Msol)]^{0.4}$, which provides a good approximation for $M_{500}\gtrsim5\times10^{13}\Msol$.}
\end{figure*}

Finally, using the parameter values from our MCMC chain, we derive observational constraints on the optical depth profiles of galaxy groups and clusters. The results are shown in Fig.\,\ref{fig:tau_prof}, for masses $M_{500}=5\times 10^{13}$, $10^{14}$, $5\times 10^{14}$, and $10^{15}\Msol$, all at $z=0.5$. In each panel, we show 50 lines with parameters determined from the MCMC chain, illustrating the modeling uncertainty in the profile. For comparison, we also show the fiducial model from \citet{Shaw:2010mn} (red dashed line). The original Shaw model does not make an attempt to model the effects of gas cooling (which causes gas to condensate into the cluster center) and therefore under-predicts the optical depth inside the core region. The differences between our new model and the original Shaw model are however small on scales that are larger than the 1\,arcmin instrument beam (black dashed line).

In Table\,\ref{table_tau_values} we summarize our constraints on $\tau(r=0)$, $\tau(r=R_{500})$, and $\tau_{500}$, i.e.\ the average $\tau$ within a disk with angular radius of $\theta_{500}$ in the plane of the sky. We find that the remaining uncertainties in the central optical depth are better than 12\% at 95\% confidence level. Current CMB experiments (with a typical beam size of 1\,arcmin) are sensitive to the gas extending out to about $R_{500}$, and the uncertainty around $R_{500}$ is considerably smaller ($\lesssim 5$\%). For $\tau_{500}$ (which is most relevant for kSZ data analyses and derived cosmological constraints) we show that \emph{the current modeling uncertainty is $\lesssim 6\%$}, depending on the cluster mass.\footnote{We constrain the $\tau$ of more massive objects better than that of low-mass objects, because higher-mass objects are generally less affected by still poorly constrained star-formation and feedback physics.}

In Fig.\,\ref{fig:tau_Delta_z} we show the prediction of our observationally calibrated ICM model for $\tau_{\Delta}$, i.e.\ the average $\tau$ within a disk of angular radius $\theta_{\Delta}$ in the plane of the sky. We choose to present our results for 3 different values that are commonly used in the literature, $\Delta=2500$, $\Delta=500$, and $\Delta=200$.

Comparing the 3 panels in Fig.\,\ref{fig:tau_Delta_z} for fixed redshift, we see that $\tau_{\Delta}$ increases with increasing $\Delta$, because higher $\Delta$ values correspond to smaller aperture sizes and hence get more weights on the high-$\tau$ regions near the center. For a fixed $\Delta$, $\tau_{\Delta}$ increases with increasing redshift, because the average density is higher at higher redshift, which leads to a higher electron number density, and thus $\tau$. Note that the $\tau_{\Delta}-M_{500}$ relation can be well approximated as a simple power-law relation for the mass range of $M_{500}\gtrsim5\times10^{13}\Msol$. For instance, for $z=0.5$ and $\Delta=500$ we find that $\tau_{500} \simeq 0.95\times10^{-3}\,[M_{500}/(10^{14}\Msol)]^{0.4}$ provides a good approximation (blue dashed line in the middle panel of Fig.\,\ref{fig:tau_Delta_z}).


\section{Discussion}

\subsection{Implication for future kSZ measurements} 

Our very strong prior on $\tau$ from our observationally calibrated ICM model can be used to break degeneracies between $\tau$ and cosmological parameters, in particular the parameter combination $f\sigma_8^2$, in future pairwise kSZ measurements. Given $\sigma_8$ from other measurements such as cluster counts, this will lead to percent-level constraints on $f$. Alternatively, the pairwise kSZ measurements can be combined with measurements of redshift space distortions from the same sample, which probe $f\sigma_8$ (e.g., \citealt{Percival:2008sh}), in order to break the degeneracy between $f$ and $\sigma_8$.

In addition, our ICM model provides tight constraints on the template of the kSZ power from groups and clusters. When combined with improved measurements of the total kSZ power spectrum with future experiments, our model can help constrain the amount of kSZ power originating from patchy reionization, which in turn shed insights into the duration and models of reionization.

Furthermore, our $\tau$-profile model can be used to design a matched filter for optimally extracting the kSZ signal from CMB data. The profile going into the matched filter does matter: \citet{Soergel:2016mce} report a reconstructed optical depth $10^3\tau$ of $3.75\pm0.89$, assuming a beta profile shape with a core radius $\theta_c=0.5^{\prime}$, but a more than twice as large amplitude ($8\pm1.82$) when assuming a projected NFW shape with $\theta_{500}=1.5^{\prime}$ instead. This demonstrates that the assumed ICM profile in the matched filter has a significant effect on the recovered kSZ amplitude. Thus, a well calibrated $\tau$-profile, such as the one presented in this work, will be critical for the accurate recovery of the kSZ signal from the upcoming surveys. 

In this work, we have calibrated the $\tau$ profile using X-ray measurements of gas density profiles of clusters for a wide redshift range ($0\lesssim z \lesssim 1.4$), and $M_{\mathrm{gas}}-M$ relations of groups and clusters covering a wide mass range ($10^{13}<M_{500}/\Msol<10^{15}$) at $z\lesssim 0.2$. Note, however, that our model is not calibrated for low-mass objects at high redshift, because requisite X-ray measurements in this range currently do not exist. Future data from the eROSITA instrument\footnote{http://www.mpe.mpg.de/eROSITA}, which will measure ICM profiles for over 100\,000 galaxy groups and clusters, will be critical for constraining our model for extending X-ray calibration of the $\tau$ profiles of high-redshift groups. 

\subsection{Residual systematic uncertainties} 
\label{sec:residual_uncertainties}

There are several residual astrophysical uncertainties in translating kSZ measurements into the pairwise velocities and hence cosmological constraints.

\paragraph{Gas Clumping} One of the systematic uncertainties in X-ray calibration of $\tau$ profiles stems from the ICM inhomogeneities associated with gas clumps. Hydrodynamical simulations suggest that gas clumping can cause the overestimate of X-ray derived gas mass by up to $16\%$ \citep{Mathiesen:1999}, if high-density clumps are not removed at all. However, high angular-resolution Chandra X-ray spectro-imaging observations can remove prominent gas clumps and reduce the ICM mass bias at the level of $\lesssim 6\%$ \citep{Nagai:2007a}. Note that the effects of gas clumping depend not only on cluster astrophysics and dynamical state \citep{Zhuravleva:2013}, but also on detailed observing conditions (such as angular resolution, source redshift, and exposure time etc). Thus, further work is needed to better quantify the clumping bias for the McDonald et al. sample, especially at high redshift. Moreover, since the effect of gas clumping is expected to increase with radius \citep{Nagai:2011,Roncarelli:2013lwa,Battaglia:2014cga}, gas clumping could become one of the major sources of systematic uncertainties in X-ray calibration of the $\tau-M$ relation in the outskirts of groups and clusters with future data.

\paragraph{Velocity substructure} Another systematic uncertainty comes from velocity substructure. Hydrodynamical simulations show that the internal velocities of the ICM could be of the same order as the overall cluster peculiar velocity. When averaging the kSZ signal inside the virial region, this velocity substructure introduces a dispersion into the signal which translates into a velocity dispersion of $50-100$\,km/s, depending on the projection of the cluster and its internal dynamical state (e.g., \citealt{Nagai:2003nw}). Cluster rotation can also be of order a few to tens of km/s (e.g., \citealt{Cooray:2002,Chluba:2002es}). Note, however, that velocity substructure does not introduce a bias into the reconstructed pairwise velocity, and is thus expected to average out when applying the pairwise estimator to a large sample of objects. A dispersion for individual objects of $\Delta v=100\,\mathrm{km/s}$ translates into an uncertainty in the mean pairwise velocity of $\Delta v / \sqrt{N} \simeq 3\,\mathrm{km/s}$ with a sample size of $N=1000$ (for comparison, \citet{Soergel:2016mce} used 6693 clusters in their analysis). This leads to only percent-level errors on pairwise velocities, which are typically of order $100\,\mathrm{km/s}$.

\paragraph{Uncertainties in the mass} The mass of a cluster is poorly known. In order to define a cluster sample for a pairwise kSZ analysis, one resorts to a proxy for the cluster mass, such as the optical richness \citep{Rykoff:2011xi}. However, there is considerable scatter in the richness-mass relation, which introduces more low-mass object compared to high-mass objects into the sample, owing to the steepness of the mass-function. Because lower-mass objects produce a smaller kSZ signal, this intrinsic scatter in the richness-mass relation introduces a bias in the pairwise kSZ amplitude, similar to the Eddington bias. F16 showed that this bias is of order 10\% (20\%), if the scatter in mass for a fixed richness is 20\% (40\%). Development of robust mass proxies is therefore another important requirement for future kSZ studies.

\paragraph{Uncertainties in the HSE bias} In this work we have assumed the model for the HSE bias from \citet{Shi:2016}, and have neglected uncertainties in that model, which can lead to additional uncertainties in $\tau$. For instance, if the uncertainty in the HSE bias is $\sim10\%$, this would lead to an additional uncertainty in $\tau_{500}$ of $\sim4\%$, given the slope of 0.4 in the $\tau_{500}-M_{500}$ relation.

\paragraph{Stellar mass-halo mass relation} Our model could be improved with external constraints on the stellar fraction inside galaxy groups and clusters, such as the ones listed in Table~\ref{tab:fstar}. However, systematic uncertainties in these measurements need to be better understood. Further improvements could be made with additional constraints on feedback in groups and clusters from observations and hydrodynamical simulations, as well as measurements of gas density profiles over a wide range of mass and redshift.

\paragraph{Mis-centering} The amplitude of the measured pairwise kSZ signal depends on cluster mis-centering, i.e.,\ the offset between the the observer-selected center and the potential minimum of the cluster. In optical data, the cluster center is assumed to be at the location of the brightest cluster galaxy (BCG). In this case, mis-centering can happen because of mis-identification of the BCG in the cluster-finding algorithm, or because the BCG is not always at the potential minimum. F16 show that this can lead to a suppression of up to $\sim 10\%$ of the overall pairwise kSZ amplitude. In order to control the astrophysical uncertainty in the kSZ cosmology to better than 10\%, it will critical improve constraints on the mis-centering distribution of the cluster sample, e.g.,\ by measuring the offset between the BCG and the SZ center (e.g., \citealt{Saro:2015lqu}).

\paragraph{Redshift errors} The pairwise kSZ amplitude further depends on the accuracy of redshift measurements, which are needed to compute the weights in the pairwise estimator in Eq.\,\ref{weights}. In a photometric survey like DES, the redshift errors of clusters are of the order $\sigma_z/(1+z)\sim0.01$ \citep{Rykoff:2016trm}, which leads to a suppression of the signal at the physical separations of order $\lesssim100$\,Mpc (F16). \citet{Soergel:2016mce} modeled the impact of redshift errors heuristically by multiplying the theoretical template with a Gaussian smoothing factor. However, a more detailed redshift error model is likely needed to realize the statistical power of future measurements. 

\paragraph{Cool-core (CC) vs.\ non-cool core (NCC) dichotomy} In this work, we have not explored the so-called CC/NCC dichotomy, i.e.,\ the fact we observe two different populations of galaxy clusters that are distinguished by having CC (high-density central regions) or NCC. The impact of the CC/NCC dichotomy is most prominent for the central optical depth. With our best-fit model we find a central optical depth of $\tau_0=4.06\times10^{-3}$ for $M_{500}=10^{14}\,\Msol$ and $z=0.5$. If we switch off cooling (i.e., set $\Gamma'=1.2$ and $\gamma=0$) in our model, we obtain $\tau_0=3.44\times10^{-3}$, i.e.\ $18\%$ lower (compared to $7\%$ modeling uncertainty), demonstrating that this is an important effect that must be included for modeling the $\tau$-profiles in the central region. Because of the redshift-evolution of cooling, this difference is smaller at higher redshift: 13\% at $z=1$ and $8\%$ at $z=1.5$. However, current and future CMB experiments are more sensitive to the integrated $\tau$; for $\tau_{\theta}$ with an aperture $\theta=1.3^{\prime}$, we find that the difference between our best-fit model with and without cooling is $<1\%$. 

\paragraph{kSZ signal from filaments}
The tSZ signal scales with the gas mass weighted temperature and hence receives a negligible contribution from regions outside halos. The kSZ signal, on the other hand, scales with the integrated electron number density and receives additional contributions from filaments and the intergalactic medium (see e.g., \citealt{AtrioBarandela:2008rq,Dolag:2015dta}). When a matched filter is applied to optimally extract the cluster kSZ component from CMB data, this additional component is expected to be negligibly small \citep{Flender:2015btu}. Note, however, that for larger filter apertures they could produce an additional bias in the measured signal.

\paragraph{kSZ signal from patchy reionization}
Another potential systematic uncertainty arises from the kSZ signal from patchy reionization, which is expected to roughly double the total kSZ power in the range $\ell=3000-10000$ \citep{Iliev:2006un}. However, since that signal is uncorrelated with the kSZ signal from groups and clusters, it is expected to average out when stacking a large number of objects.

\vspace{2mm}
Addressing these remaining uncertainties above will further improve cosmological constraints based on pairwise kSZ measurements.


\section{Conclusions}

The pairwise kSZ signal has emerged as a new, powerful probe of cosmology and gravity. However, the power of kSZ cosmology is currently limited by the uncertainty in the optical depth of galaxy groups and clusters. In this work, we have derived observational constraints on the optical depth of galaxy groups and clusters, by developing a physically motivated, computationally efficient semi-analytical model of the ICM and constraining it using the state-of-the-art X-ray observations of galaxy groups and clusters. Our main results are summarized as follows:

\bim

\item We have presented a new model for the ICM, which takes into account star-formation, feedback, non-thermal pressure, \emph{and} gas cooling, which is modeled as a change in the effective EOS in the central regions. Note that the effects of gas cooling were not modeled in the earlier work by \citet{Shaw:2010mn}. This additional feature is critical for describing the observed gas density profiles of galaxy clusters and constraining the external prior on the optical depth of groups and clusters. 

\item Our semi-analytic model is computationally efficient and can reproduce the recent results from hydrodynamical simulations presented in \citet{Battaglia:2016xbi}. Our best-fit model is consistent with the results of recent hydrodynamical simulations that include a variety of cluster astrophysics, including gas cooling, star formation, and energy feedback from AGN/SNe.

\item We have calibrated the ICM model using the recent X-ray data, including measurements of gas density profiles of massive clusters \citep{McDonald:2013fka} as well as the $M_{\mathrm{gas}}-M$ relation from groups and clusters \citep{0004-637X-640-2-691,Sun:2008eh,2015A&A...573A.118L}. These observations provide powerful constrains on the physically-motivated parameters of the model (including gas cooling, star formation, and energy feedback from AGN/SN) over cosmic time. 

\item Most importantly, \emph{our observationally calibrated model predicts the average, integrated $\tau$ to better than 6\% modeling uncertainty} (at 95\% confidence level), indicating that the uncertainty associated with the ICM modeling is no longer a limiting factor. 

\item The remaining uncertainties in the optical depth are selection effects and astrophysical uncertainties described in \S\ref{sec:residual_uncertainties}.
If these additional uncertainties can be better understood, our model for the optical depth should break the degeneracy between optical depth and cluster velocity in the analysis of future pairwise kSZ measurements and improve cosmological constrains from the combination of upcoming galaxy and CMB surveys, including the nature of dark energy, modified gravity, and neutrino mass. 
\eim

Further advances in our understanding of the structure and evolution of galaxy groups and clusters will help maximize the scientific return from the upcoming galaxy and CMB surveys. 

\section*{Acknowledgements}
We thank the organizers of the SnowPAC 2016 conference, where this work was initiated. We also acknowledge 
Nick Battaglia, 
Bradford Benson, 
Sebastian Bocquet, 
Francesco De Bernardis,
Simone Ferraro, 
Salman Habib, 
Erwin Lau, 
Emmanuel Schaan, 
Bjoern Soergel,
David Spergel,
Naonori Sugiyama,
Kyle Story, 
and the anonymous referee
for useful discussions and/or comments on the manuscript. 
We acknowledge use of the software packages pyGTC\footnote[1]{\texttt{http://pygtc.readthedocs.io}} \citep{Bocquet2016} and Emcee\footnote[2]{\texttt{http://dan.iel.fm/emcee/current/}} \citep{2013PASP..125..306F}. 
This work is supported by NSF grant AST-1412768, and NASA GO4-15122A and GO5-16141X. 
Argonne National Laboratory's work was supported under the U.S.\ Department of Energy contract DE-AC02-06CH11357. 
This research used resources of the National Energy Research Scientific Computing Center, a DOE Office of Science User Facility supported by the Office of Science of the U.S.\ Department of Energy under Contract No.\ DE-AC02-05CH11231.

\bibliography{bib.bib}
\end{document}